\documentstyle[namedreferences,proceedings]{crckapb}

\newcommand{\gtrapprox}{\stackrel{\scriptscriptstyle>}{\scriptscriptstyle\sim}}

\begin{opening}
\title{Separating intrinsic and microlensing \protect\\
  variability using parallax measurements}
\author{Stein Vidar Hagfors Haugan}
\institute{Institute of Theoretical Astrophysics, University of Oslo\\
  Pb. 1029, Blindern\\
  N-0315 OSLO\\
  {\tt http://www.uio.no/\~\/steinhh/index.html}}
\end{opening}

\runningtitle{}

\newcommand{\A}{{\rm\bf A}}
\newcommand{\B}{{\rm\bf B}}
\newcommand{\I}{{\rm\bf I}}
\newcommand{\FA}{F_\A}
\newcommand{\FB}{F_\B}
\newcommand{\FI}{F_\I}
\newcommand{\FIA}{F_{\I\A}}
\newcommand{\muA}{{\mu_\A\!}}

\newcommand{\deltat}{\Delta t}
\newcommand{\DAB}{D_{\A\B}}
\newcommand{\vperp}{v_\perp}
\newcommand{\ti}{t_i}
\newcommand{\filter}[2]{{\Phi\!\left[#2;#1\right]}}
\newcommand{\taui}{{\tau_\I}}
\newcommand{\taumu}{{\tau_\mu}}
\newcommand{\AI}{A_{\I}}
\newcommand{\Amu}{A_{\mu}}

\begin{document}

\section{Introduction}
In gravitational lens systems with 3 or more resolved images of a
quasar, the intrinsic variability may be unambiguously separated from
the microlensing variability through parallax measurements from 3
observers when there is no relative motion of the lens masses
\cite{Refsdal93}. In systems with fewer than 3 resolved images,
however, this separation is not straightforward.  For the purpose of
illustration, I make the following simplifications for the
one-dimensional case: The observations consist of well-sampled time
series of the observed flux $\FA(\ti)$ and $\FB(\ti)$ at two points in
the observer plane.  The separation vector of the two points is
parallel to the direction of the transversal motion of the
source-lens-observer system, and the distance $\DAB$ between the
observers is known.  Furthermore, the distance $\DAB$ is small
compared to the typical length scale of fluctuations in the
magnification $\mu(x)$.

It is possible to calculate the ratio of the instantaneous
magnification at the two observers as a function of time, defined by
\vspace{-.1cm}
\begin{equation}
  r(\ti) = \FB(\ti)/\FA(\ti)
  \label{eq:ratio-history}
  \vspace{-.15cm}
\end{equation}
where $\FA(\ti)$ and $\FB(\ti)$ are the observed fluxes at observer
\A\  and \B\ respectively. I am assuming that observer \B\ is the
leading one.

With these assumptions, the magnification history $\muA(\ti)$ for
observer \A, can be reconstructed (apart from boundary conditions)
through the formula
\vspace{-.2cm}
\begin{equation}
  \muA(\ti) = \muA(\ti - \deltat) r(\ti - \deltat) %
  \;\;\;\;\;{\rm with}\;\;\;\;\; %
  \deltat = \frac{\DAB}{\vperp}
  \label{eq:muA-formula}
\end{equation}
where $\vperp$ is the unknown velocity perpendicular to the line of
sight.

Given a velocity $\vperp$, the microlensing magnification history
$\muA$ is uniquely determined, and thereby also the intrinsic flux,
given by
\vspace{-.31cm}

\begin{equation}
  \FIA(\ti) = \FA(\ti)/\muA(\ti)
  \label{eq:intrinsic-history}
\end{equation}

The velocity is chosen by minimizing some measure of the
variability (e.g.,~$\chi^2$) of $\FIA$, given by
\(
  \chi^2 = \sum_{i=1}^N (\FIA(\ti)-\langle\FIA\rangle)^2
  \label{eq:chisquare}
\)

\section{Preliminary results}
\label{sec:results}

In order to test the method, dummy data for the intrinsic flux
$\FI(\ti)$ and the magnification $\mu(x_i) = \mu(\vperp \ti)$ were
made by simply filtering white noise, $N(t)$, with gaussian low-pass
filters with characteristic scales $\taui$ and~$\taumu$, and then
exponentiating, e.g.:
\begin{equation}
  \begin{array}[c]{lcl}
    \FI(t)  & = &  \exp(\AI\filter{\taui}{N(t)})\\
    \mu(t)  & = &  \exp(\Amu\filter{\taumu}{N(t)})
  \end{array}
  \label{eq:exp-filtered-noise}
\end{equation}
where $\filter{\tau}{\ldots}$ denotes gaussian filtering with time
scale~$\tau$, and then renormalization to make the variance equal to
one. $\AI$ and $\Amu$ are the amplitudes of the intrinsic and
microlensing variabilities, respectively. For simplicity, but without
loss of generality, the units were chosen so that the ``true''
source-lens-observer transversal velocity~$\vperp$ and the
characteristic scale of the magnification fluctuations $\taui$ were
equal to~1.  The observations were simulated according to

\vspace{-.2cm}
\begin{equation}
  \begin{array}[c]{lcl}
    \FA(\ti) & = & \FI(\ti) \; \mu(\vperp \ti)\\[0cm] %
    \FB(\ti) & = & \FI(\ti) \; %
    {}              \mu\!\left(\vperp \ti + %
    {}                    \frac{\DAB}{\vperp}\right) %
  \end{array}
  \label{eq:observations}
  \vspace{-.1cm}
\end{equation}

The flux ratio $r(\ti)$, the magnification history $\muA(\ti)$ and the
intrinsic flux $\FIA(\ti)$ were calculated for a range of values for
$\vperp$.  For a wide range of parameters, the $\chi^2$ function is
fairly well-behaved, with a quadratic minimum, although the minimum
may be somewhat displaced compared to the true value of $\vperp$. The
most difficult cases seem to be those where $\taui\!\approx\!\taumu$
and $\AI\!\gtrapprox\!\Amu$.

It is unclear how useful this method is for the two-dimensional case
with two observers. This will be the subject of further study.  The
extension of the method to 3 observers in two dimensions with is fairly
straightforward.  In cases where relative motion of the lensing point
masses are important, only a partial separation will be possible.


\begin{thebibliography}{}

\bibitem[\protect\citeauthoryear{Refsdal}{1993}]{Refsdal93}
Refsdal, S. 1993, in {\it Gravitational Lenses in the Universe},
eds. Surdej et al., Universit{\'e} de Li{\`e}ge, Belgium
\end{thebibliography}
\end{document}